\documentclass[10pt, conference]{IEEEtran}
\IEEEoverridecommandlockouts
\usepackage{cite}%[nocompress]{cite}
\usepackage{amsmath,amssymb,amsfonts}
\usepackage{comment}
\usepackage{algorithmic}
\usepackage{graphicx}
\usepackage{textcomp}
\usepackage{xcolor}
\usepackage[pscoord]{eso-pic}
\usepackage{balance}
\usepackage{fancyvrb}

\def\BibTeX{{\rm B\kern-.05em{\sc i\kern-.025em b}\kern-.08em
    T\kern-.1667em\lower.7ex\hbox{E}\kern-.125emX}}

\def\BibTeX{{\rm B\kern-.05em{\sc i\kern-.025em b}\kern-.08em
    T\kern-.1667em\lower.7ex\hbox{E}\kern-.125emX}}
\def\BibTeX{{\rm B\kern-.05em{\sc i\kern-.025em b}\kern-.08em
    T\kern-.1667em\lower.7ex\hbox{E}\kern-.125emX}}

\usepackage{pifont}
\usepackage{framed}
\usepackage{soul}
\usepackage{booktabs}  
\usepackage{csquotes}
\usepackage{multirow}
\usepackage{array}
\newcolumntype{L}[1]{>{\raggedright\let\newline\\\arraybackslash\hspace{0pt}}m{#1}}
\newcolumntype{C}[1]{>{\centering\let\newline\\\arraybackslash\hspace{0pt}}m{#1}}
\newcolumntype{R}[1]{>{\raggedleft\let\newline\\\arraybackslash\hspace{0pt}}m{#1}}
\usepackage{url}
\MakeOuterQuote{"}
\usepackage[caption=false,font=footnotesize,labelformat=simple]{subfig} % caption=false is needed for IEEE template
\usepackage[bookmarks=false,hidelinks]{hyperref}

\usepackage{acronym}
\acrodef{IP}[IP]{intellectual property block}
\acrodef{SoC}[SoC]{System-on-Chip}
\acrodef{IC}[IC]{integrated circuit}
\acrodef{eFPGA}[eFPGA]{embedded field programmable gate array}
\acrodef{RTL}[RTL]{register-transfer level}
\acrodef{CPS}{Cyber-Physical System}
\acrodef{IoT}{Internet of Things}
\acrodef{CAD}{Computer-Aided Design}
\acrodef{EDA}{Electronic Design Automation}
\acrodef{HPC}{High-Performance Computing}
\acrodef{DL}{deep learning}
\acrodef{ML}{machine learning}
\acrodef{NLP}{natural language processing}
\acrodef{IC}{Integrated Circuit}
\acrodef{CWE}[CWE]{Common Weakness Enumeration}
\acrodef{CVE}[CVE]{Common Vulnerabilities and Exposures}
\acrodef{LLM}[LLM]{large language model}
\acrodef{NMT}[NMT]{neural machine translation}
\acrodef{IP}[IP]{hardware intellectual property block}
\acrodef{HDL}[HDL]{hardware description language}
\acrodef{RTL}[RTL]{register-transfer level}
\acrodef{SDL}[SDL]{security development lifecycle}
\acrodef{FSM}[FSM]{finite state machine}
\acrodef{AST}[AST]{abstract syntax tree}
\acrodef{SoC}[SoC]{system-on-chip}

\begin{document}
\bstctlcite{IEEEexample:BSTcontrol}
\IEEEtriggercmd{\balance}
\IEEEtriggeratref{14}

\title{%
    The Quest to Build Trust Earlier in Digital Design %
}

\author{%
    \IEEEauthorblockN{Benjamin Tan}
    \IEEEauthorblockA{\textit{Department of Electrical and Software Engineering} \\
        \textit{University of Calgary}\\ 
        \textit{Calgary, Alberta, Canada}\\
        benjamin.tan1@ucalgary.ca%
        \thanks{
            The research program that this work describes is supported in part by the Natural Sciences and Engineering Research Council of Canada (NSERC) [RGPIN-2022-03027]. Cette recherche a été financée en partie par le Conseil de recherches en sciences naturelles et en génie du Canada (CRSNG). 
            The research is also supported in part by Alberta Innovates and the University of Calgary, and by a gift from Intel Corporation. This work does not in any way constitute an Intel endorsement of a product or supplier.
        }
    }
}

\IEEEtitleabstractindextext{
\begin{abstract}
The ever-rising complexity of computer systems presents challenges for maintaining security and trust throughout their lifetime. 
As hardware forms the foundation of a secure system, we need tools and techniques that support computer hardware engineers to improve trust and help them address security concerns. 
This paper highlights a vision for tools and techniques to enhance the security of digital hardware in earlier stages of the digital design process, especially during design with hardware description languages. 
We discuss the challenges that design teams face and explore some recent literature on understanding, identifying, and mitigating hardware security weaknesses as early as possible. 
We highlight the opportunities that emerge with open-source hardware development and sketch some open questions that guide ongoing research in this domain.

\end{abstract}
}

\maketitle
\IEEEdisplaynontitleabstractindextext

\section{Introduction\label{sec:intro}}

Designing computer systems is challenging.
Not only do designers have to work hard to satisfy functional requirements (often under considerable time pressure), but increasing device interconnectivity and desire for computers in sensitive applications introduce security requirements into the fold. 
Naturally, we want to identify potential shortcomings in security in earlier stages of digital design, thus building trust in our overall system. 
Building trust earlier in design by identifying and addressing potential weaknesses is also beneficial because as we progress through the design process, the cost of design changes considerably increases. 
As best practice, designers should consider adopting a security development lifecycle (SDL) (e.g.,~\cite{dorsey_intel_2020}) where a security mindset is adopted throughout the design process. 
Teams need to define security objectives, formulate meaningful threat models, implement, and then verify and validate security mechanisms. 
Careful thought about support over the lifespan of a released product in the field is needed. 

However, while software designers have at their disposal many potential tools to help with security throughout the design flow (e.g.,~\cite{owasp_source_nodate}), hardware designers do not yet have such luxury~\cite{bidmeshki_hunting_2021,Dessouky_hardfails_2019,Ahmad_cweat_2022}. 
In fact, systemization of how we think about security weaknesses is emerging and evolving, with recent efforts like the introduction of the hardware \acp{CWE}~\cite{the_mitre_corporation_cwe_2022} and standardization efforts like Accellera's Security Annotation for Electronic Design Integration Standard (SA-EDI)~\cite{accellera_systems_initiative_security_2021} and IEEE's P3164 working group~\cite{ieee_p3164_working_group_p3164_nodate} revealing industry-led efforts to tackle security issues. 
Recent competitions like Hack@DAC~\cite{Dessouky_hardfails_2019,noauthor_home_2024} seek to raise awareness and engagement with security bugs. 

Even so, there remains a gap between the accessibility of (hardware) cybersecurity expertise and the need for secure design. 
As hardware forms the foundation of a secure system, we need tools and techniques that support computer hardware engineers to improve trust and help them address security concerns. 
\textit{How do we choose what security features to implement? 
How do we check our designs, even if \underline{designs are not yet complete}? 
How do we build trust, even if designers are not security experts?}
Such questions are not easily solved. 

Towards the goal of building trust in digital systems, this paper highlights a vision for tools and techniques to enhance the security of digital hardware in earlier stages of the digital design process and some of the progress our team has made in this quest for improving security. 
We discuss the challenges that design teams face and explore some recent literature on understanding, identifying, and mitigating hardware security weaknesses as early as possible. 
With the emphatic growth in open-source hardware design, there is an opportunity to learn from and contribute to open-source ecosystems in pushing our understanding and handling of security challenges.

The rest of this paper is as follows. \autoref{sec:bg} provides background on the area of hardware security bugs and the motivation for wanting tools and techniques to support things earlier in design. 
In~\autoref{sec:dirs}, we discuss some of our recent work in the area and highlight some open challenges, and present some related work in~\autoref{sec:rw}. 
\autoref{sec:conc} concludes.

\section{Background and Motivation\label{sec:bg}}

Hardware security is a wide and varied field, and our understanding of risks continues to evolve. 
There are potential security issues such as threats in the supply chain (especially given globalized production~\cite{rostami_primer_2014}), the potential for malicious modifications~\cite{xiao_hardware_2016}, or even perhaps unintentional bugs~\cite{Dessouky_hardfails_2019}. 
Others include issues that can manifest physically (e.g., side-channels~\cite{standaert_introduction_2010}). 
There are many potential solutions for different security challenges such as new mechanisms~\cite{tan_challenges_2022} and the domain features back-and-forth developments, likened to ``cat-and-mouse'' games for attacks and defenses (e.g., in logic locking~\cite{chakraborty_keynote_2019}).
Naturally, to build trusted systems, we want to choose and combine the ``best'' security solutions available. 
However, what is ``best'' is determined on a case-by-case basis; designers might need to favor one design metric over another, such as keeping area overhead low, maximizing performance, or improving usability—sometimes at the cost of security. 

Given the plethora of options available, security-driven design remains largely a human-in-the-loop endeavor. 
Choosing how best to go about improving trust requires creativity and value-based judgment. 
Even the process of deciding what is important (i.e., identifying \textit{assets}) is subjective and requires some level of security expertise and a handle on designer intent~\cite{ieee_p3164_working_group_asset_2024}. 
However, when there are humans involved, there is the risk of unintentional mistakes -- in other words, there is a potential for \textit{bugs}, a design defect that might result in unintended behavior (noting, of course, that designer intent is often imperfect, incomplete, or implicit~\cite{widder_what_2024}). 
Design bugs can appear throughout the design lifecycle, ranging from improperly captured (or even defined) specifications to literal typographical errors in the code. 

\begin{figure}[t]
    \centering
    \includegraphics[width=0.99\columnwidth]{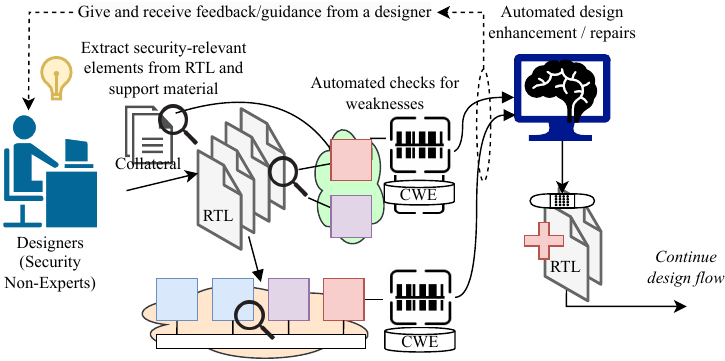}
    \caption{A vision for tools and techniques to help with building trust in early stages of design}
    \label{fig:vision1}
\end{figure}

\textbf{How can we effectively find and deal with bugs?} This question motivates the ``quest'' outlined in this paper.
While one can (and should) think about security throughout the lifecycle, our team's work is especially concerned with what is possible at \textit{early} stages of design, i.e., during RTL implementation \underline{and earlier}. 
\textit{Why?} If we find bugs early in design, we can make lower-cost changes to reduce risk and avoid calamities should an exploitable vulnerability ``escape'' in a final product. 
Working at RTL and earlier improves the likelihood that we have at our disposal more markers of designer intent, ranging from code comments, signal names, specification documents, and such -- we can work \textbf{with} designers to give guidance (and hopefully improve security awareness) as well as receive guidance (for example, on the validity of an identified weakness). 
How early do we envision? As early as possible, potentially even before we finalize security objectives and threat models -- this entails adaptable and flexible analyses. 
As of today, there is no panacea when it comes to identifying security weaknesses~\cite{Ahmad_cweat_2022,Dessouky_hardfails_2019}. 
\autoref{fig:vision1} illustrates a vision for tools and techniques that can assist designers in building trust. 

Recent efforts like the hardware \acp{CWE}~\cite{the_mitre_corporation_cwe_2022} provide a means to categorize and reason over security weaknesses in a common framework, but how one can use the framework, e.g., for implementing automated detection tools or informing designers, remains an open challenge. 
Another industry-led effort, the SA-EDI standard~\cite{accellera_systems_initiative_security_2021} (now transitioned to an in-progress IEEE proposed standard, IEEE P3164), captures the idea that security is a collective responsibility by aiming for a standardized format for security-related collateral associated with an IP -- a system \textit{integrator} can use the information to make an ``informed decision at the time of IP integration'' and lead to actions like ``implement[ing] mitigations'' or even deciding that 
the risks are out of scope''~\cite{accellera_systems_initiative_security_2021}. 
To the best of our knowledge, uptake of the standard has been low, perhaps partly due to the onerousness of populating the fields required (by hand). 
\textit{Can we automate things? If so, how much automation should we have? }
In light of everything discussed so far, some of the open questions include (and are not limited to):
\begin{itemize}
    \item \textbf{Bugs, bugs, bugs}: How can we detect different kinds of security weaknesses and bugs? Are there some types of bugs that are inherently more (or less) amenable to certain types of analyses? 
    \item \textbf{Humans-in-the-loop}: How can we maximize human expertise and intent in building trust? Is there a space between the extremes of having everything manually crafted and fully automated? How do we support the design processes currently used by design teams?
\end{itemize}

\section{Recent Directions\label{sec:dirs}}
    Given our discussion of the motivation for our work, this section provides an overview of some of the directions we have been pursuing, emphasizing what we consider to be some open challenges. 
    We also provide interested readers with a non-extensive overview of related work. 

\subsection{Static Analysis}
    In pursuing security analysis at earlier design stages, we looked at \textit{static analysis}~\cite{Ahmad_cweat_2022}. 
    Static analysis focuses primarily (if not solely) on source code, thus obviating the need for other design collateral (such as testbenches and a functioning simulation environment). 
    In some cases, analyzing \textit{incomplete} code is even possible. 
    In that work, we identified five \acp{CWE}, \textit{CWE-1234}: Hardware Internal or Debug Modes Allow Override of Locks, \textit{CWE-1271}: Uninitialized Value on Reset for Registers with Security Settings, \textit{CWE-1245}: Improper Finite State Machines (\acp{FSM}) in Hardware Logic, \textit{CWE-1280}: Access Control Check Implemented After Asset is Accessed and \textit{CWE-1262}: Improper Access Control for Register Interface, as amenable to pattern recognition in a ``context-less'' fashion. 
    This means that we could craft scanners to identify potential instances of these weaknesses without requiring additional context from a designer, such as specific assets or design intent, as these weaknesses could be considered ``general''. 
    
    \texttt{CWEAT}~\cite{Ahmad_cweat_2022} showed that certain \acp{CWE} can be detected during the early RTL Implementation, where we were able to highlight 180 instances of potential weaknesses, reducing the search space for manual checking. 
    However, while promising, several challenges emerged -- as with any imperfect detection system, there is the risk of false positives that can distract, confuse, or burden a human operator. 
    As such, more work is needed to reduce the ``noise.''
    We did, however, find that similar scanning could be used in cases where the RTL code is more structured, such as that generated through HLS~\cite{collini_using_2023}. 
    In investigating the \acp{CWE}, we found that several entries are quite ``broad'' -- given that we only looked at five of (as of writing) 108 hardware-relevant \acp{CWE}, we surmised that the remaining require much more \textit{context}. 
    In other words, scanners need to incorporate design- or project-specific information to guide the identification of areas of concern. 
    As of now, we lack robust solutions for context-inclusive scanning. 
    
\subsection{Large Language Models (LLMs)}
    The recent emergence of \acp{LLM} has inspired a flurry of research activity (readers might find the recent survey~\cite{qiu_explaining_2024} useful). 
    Keeping the focus on \textit{source code}, we have investigated LLMs for detecting potential security bugs~\cite{ahmad_flag_2023}, repairing bugs~\cite{ahmad_hardware_2024}, and assertion generation~\cite{kande_security_2024}. 
    In many ways, these works could be considered a type of ``static analysis,'' as previously discussed, at least when using LLMs without ``feedback'' (such as from a simulation or formal verification tool). 
    
    Our proof-of-concept implementations\footnote{e.g., \url{https://zenodo.org/records/10416865}} show that there is the potential for using models to help designers, although sometimes with numerous requests to the LLMs. 
    We speculate that their usefulness will increase over time, at least while new LLM models continue to exhibit increased performance in general. 
    However, like other LLM-centric solutions, LLM shortcomings remain, such as hallucination or mixed-quality in their outputs (such as security~\cite{pearce_asleep_2022}). 
    Prompt engineering and verification/evaluation of LLM outputs are some of the challenges we continue to face. 
    As of now, we postulate that the more "interactive" nature of LLMs (e.g., for "chat") can provide opportunities to more directly inform designers or act as an interface for designer guidance. 
    For example, our recent work investigated using LLMs to explain EDA tool error messages to novice users~\cite{qiu_explaining_2024}.

\subsection{Learning from Open Source Processes}
    With the rise in open-source hardware projects (e.g., OpenTitan~\cite{opentitan_github_2024}) and heterogeneous SoCs, there is an opportunity to learn from hardware development. 
    This is especially pertinent from an academic point of view, given that access to internal details of commercial designs is understandably unlikely. 
    Having access to designs written in HDLs is very useful for experimentation and analysis, and efforts like Hack@DAC (part of the Hack The Silicon series~\cite{noauthor_home_2024}), TrustHub~\cite{noauthor_trust-huborg_nodate}, CAD4Security~\cite{noauthor_cad4security_nodate} and CAD for Assurance~\cite{noauthor_cad_nodate} have enabled scores of research, including ours. 
    Even so, the quantity of open-source hardware remains orders of magnitude less than software, with datasets of known bugs even more scarce. 

    What we do have available, however, can be very interesting. 
    In our recent work~\cite{ah-kiow_investigation_2024}, we have begun to look beyond the RTL code and into the \textit{discussions} accompanying digital design, where human developers identify potential bugs, discuss them, and implement fixes. 
    By looking at behaviors, such as the nature of identified issues, the magnitude of code changes between commits, and the levels of discussion, we can start to build a more holistic view of the development process. 
    When we manually examined OpenTitan, we found that 53\% of the bugs identified during its development (in the period that we examined) had potential security implications and that 55\% of all bug fixes changed only a single file. 
    We think that more consideration of open-source projects can reveal new insights, and as more projects emerge and activity increases (and we hope they do), this opportunity will grow.

\section{Selected Related Work\label{sec:rw}}
Finding hardware security bugs in the design stage at RTL requires considerable security expertise, especially for manual analysis~\cite{Dessouky_hardfails_2019,fischer_hardware_2021, bidmeshki_hunting_2021}. 
There exist some specialized approaches that require experts to devise information flow properties for formal verification and simulation~\cite{noauthor_cycuity_nodate}. 
Security invariants are mined in~\cite{zhang_end--end_2018}, and testing approaches like  concolic testing (e.g.,~\cite{meng_rtl-contest_2021}) or fuzzing (e.g.,~\cite{laeufer_rfuzz_2018}) are emerging. 
There are a few approaches for security analysis during RTL design, e.g., the construction and analysis of hyperflow graphs~\cite{meza_information_2023}, and progress toward the automation of various security tasks, such as asset identification~\cite{farzana_saif_2021} and security property re-use~\cite{zhang_transys_2020}. 
Notably, few works attempt to deal with security feedback as you go; the automated verification environment for HDLs is far less mature compared with the state-of-the-art for higher-level computer programming (which has several security-focused static analysis tools---e.g., nearly 100 listed on OWASP~\cite{owasp_source_nodate}). 
Several tools provide linting capabilities for RTL (e.g.,~\cite{noauthor_verilator_nodate}), but these do not yet focus on highlighting security weaknesses. 
As previously discussed, LLMs provide opportunities for early-stage security assistance~\cite{wang_llms_2024}, such as property generation~\cite{meng_unlocking_2023}. 

Related literature deals with hardware Trojans (HTs) detection~\cite{xiao_hardware_2016}; if we consider bugs to be ``unintentional'' artifacts, HTs are complementary ``intentional'' malicious insertions. 
There are techniques for HT detection (e.g.,~\cite{trippel_bomberman_2021,han_hardware_2019,waksman_fanci_2013}) that attempt to localize suspicious design parts using heuristics or ML techniques. 
While they can serve as a starting point for security big detection, they do not usually apply to earlier design stages or propose repair techniques for security weaknesses. While research into automatic program repair in software engineering is mature~\cite{gazzola_automatic_2019}, similar efforts for hardware design lag, but recent work is promising~\cite{Ahmad_cirfix_2022, ahmad_hardware_2024, laeufer_rtl-repair_2024}. 

\section{Conclusions\label{sec:conc}}
We gave insights into our vision for tools and techniques to enhance the security of digital hardware in earlier stages of the digital design process. 
We discussed recent literature on understanding, identifying, and mitigating hardware security weaknesses as early as possible and outlined some ongoing challenges and opportunities, especially those that continue to emerge with open-source hardware development.

\bibliographystyle{IEEEtran}
\bibliography{IEEEabrv,trefs}

\begin{thebibliography}{10}
\providecommand{\url}[1]{#1}
\csname url@samestyle\endcsname
\providecommand{\newblock}{\relax}
\providecommand{\bibinfo}[2]{#2}
\providecommand{\BIBentrySTDinterwordspacing}{\spaceskip=0pt\relax}
\providecommand{\BIBentryALTinterwordstretchfactor}{4}
\providecommand{\BIBentryALTinterwordspacing}{\spaceskip=\fontdimen2\font plus
\BIBentryALTinterwordstretchfactor\fontdimen3\font minus \fontdimen4\font\relax}
\providecommand{\BIBforeignlanguage}[2]{{%
\expandafter\ifx\csname l@#1\endcsname\relax
\typeout{** WARNING: IEEEtran.bst: No hyphenation pattern has been}%
\typeout{** loaded for the language `#1'. Using the pattern for}%
\typeout{** the default language instead.}%
\else
\language=\csname l@#1\endcsname
\fi
#2}}
\providecommand{\BIBdecl}{\relax}
\BIBdecl

\bibitem{dorsey_intel_2020}
\BIBentryALTinterwordspacing
V.~Dorsey and C.~Morhardt, ``Intel {Security} {Development} {Lifecycle},'' Intel Corporation, Tech. Rep., 2020. [Online]. Available: \url{https://newsroom.intel.com/wp-content/uploads/sites/11/2020/10/sdl-2020-whitepaper.pdf}
\BIBentrySTDinterwordspacing

\bibitem{owasp_source_nodate}
\BIBentryALTinterwordspacing
{OWASP}, ``\BIBforeignlanguage{en}{Source {Code} {Analysis} {Tools} {\textbar} {OWASP} {Foundation}}.'' [Online]. Available: \url{https://owasp.org/www-community/Source_Code_Analysis_Tools}
\BIBentrySTDinterwordspacing

\bibitem{bidmeshki_hunting_2021}
\BIBentryALTinterwordspacing
M.~M. Bidmeshki \emph{et~al.}, ``Hunting {Security} {Bugs} in {SoC} {Designs}: {Lessons} {Learned},'' \emph{IEEE Design \& Test}, vol.~38, no.~1, pp. 22--29, Feb. 2021. [Online]. Available: \url{https://ieeexplore.ieee.org/document/9154739/}
\BIBentrySTDinterwordspacing

\bibitem{Dessouky_hardfails_2019}
\BIBentryALTinterwordspacing
G.~Dessouky \emph{et~al.}, ``{HardFails}: Insights into {Software-Exploitable} hardware bugs,'' in \emph{USENIX Security Symp.}\hskip 1em plus 0.5em minus 0.4em\relax USENIX Association, Aug. 2019, pp. 213--230. [Online]. Available: \url{https://www.usenix.org/conference/usenixsecurity19/presentation/dessouky}
\BIBentrySTDinterwordspacing

\bibitem{Ahmad_cweat_2022}
\BIBentryALTinterwordspacing
B.~Ahmad \emph{et~al.}, ``{Don't CWEAT It: Toward CWE Analysis Techniques in Early Stages of Hardware Design},'' in \emph{IEEE/ACM Int. Conf. on CAD}, Dec 2022, p. 1–9. [Online]. Available: \url{https://dl.acm.org/doi/10.1145/3508352.3549369}
\BIBentrySTDinterwordspacing

\bibitem{the_mitre_corporation_cwe_2022}
\BIBentryALTinterwordspacing
{The MITRE Corporation}, ``{CWE} - {CWE}-1194: {Hardware} {Design} (4.1),'' \url{https://cwe.mitre.org/data/definitions/1194.html}, 2022. [Online]. Available: \url{https://cwe.mitre.org/data/definitions/1194.html}
\BIBentrySTDinterwordspacing

\bibitem{accellera_systems_initiative_security_2021}
\BIBentryALTinterwordspacing
{Accellera Systems Initiative}, ``Security {Annotation} for {Electronic} {Design} {Integration} {Standard},'' Jul. 2021. [Online]. Available: \url{https://www.accellera.org/images/downloads/standards/Accellera_SA-EDI_Standard_v10.pdf}
\BIBentrySTDinterwordspacing

\bibitem{ieee_p3164_working_group_p3164_nodate}
\BIBentryALTinterwordspacing
{IEEE P3164 Working Group}, ``\BIBforeignlanguage{en}{P3164 {Standard} for {Security} {Annotation} for {Electronic} {Design} {Integration}}.'' [Online]. Available: \url{https://standards.ieee.org/ieee/3164/11106/}
\BIBentrySTDinterwordspacing

\bibitem{noauthor_home_2024}
\BIBentryALTinterwordspacing
``\BIBforeignlanguage{en-US}{Home - {Hack} {The} {Silicon}},'' 2024. [Online]. Available: \url{https://hackthesilicon.com/}
\BIBentrySTDinterwordspacing

\bibitem{rostami_primer_2014}
\BIBentryALTinterwordspacing
M.~Rostami, F.~Koushanfar, and R.~Karri, ``A {Primer} on {Hardware} {Security}: {Models}, {Methods}, and {Metrics},'' \emph{Proc. {IEEE}}, vol. 102, no.~8, pp. 1283--1295, Aug. 2014. [Online]. Available: \url{http://ieeexplore.ieee.org/document/6860363/}
\BIBentrySTDinterwordspacing

\bibitem{xiao_hardware_2016}
\BIBentryALTinterwordspacing
K.~Xiao \emph{et~al.}, ``Hardware {Trojans}: {Lessons} {Learned} after {One} {Decade} of {Research},'' \emph{ACM Transactions on Design Automation of Electronic Systems (TODAES)}, vol.~22, no.~1, pp. 6:1--6:23, May 2016. [Online]. Available: \url{http://doi.org/10.1145/2906147}
\BIBentrySTDinterwordspacing

\bibitem{standaert_introduction_2010}
\BIBentryALTinterwordspacing
F.-X. Standaert, ``\BIBforeignlanguage{en}{Introduction to {Side}-{Channel} {Attacks}},'' in \emph{\BIBforeignlanguage{en}{Secure {Integrated} {Circuits} and {Systems}}}, I.~M. Verbauwhede, Ed.\hskip 1em plus 0.5em minus 0.4em\relax Boston, MA: Springer US, 2010, pp. 27--42. [Online]. Available: \url{https://doi.org/10.1007/978-0-387-71829-3_2}
\BIBentrySTDinterwordspacing

\bibitem{tan_challenges_2022}
\BIBentryALTinterwordspacing
B.~Tan, ``Challenges and {Opportunities} for {Hardware}-{Assisted} {Security} {Improvements} in the {Field},'' in \emph{2022 23rd {Int.} {Symp. } on {Quality} {Electronic} {Design} ({ISQED})}, Apr. 2022, pp. 90--95, iSSN: 1948-3295. [Online]. Available: \url{https://ieeexplore.ieee.org/abstract/document/9806254}
\BIBentrySTDinterwordspacing

\bibitem{chakraborty_keynote_2019}
A.~Chakraborty \emph{et~al.}, ``Keynote: {A} {Disquisition} on {Logic} {Locking},'' \emph{{IEEE} Trans. Comput.-Aided Design Integr. Circuits Syst.}, pp. 1--1, 2019.

\bibitem{ieee_p3164_working_group_asset_2024}
\BIBentryALTinterwordspacing
{IEEE P3164 Working Group}, ``Asset {Identification} for {Electronic} {Design} {IP},'' \emph{Asset Identification for Electronic Design IP}, pp. 1--26, Apr. 2024. [Online]. Available: \url{https://ieeexplore.ieee.org/document/10496567}
\BIBentrySTDinterwordspacing

\bibitem{widder_what_2024}
\BIBentryALTinterwordspacing
D.~G. Widder and C.~L. Goues, ``What is a "bug"? subjectivity, epistemic power, and implications for software research,'' Feb. 2024, arXiv:2402.08165 [cs]. [Online]. Available: \url{http://arxiv.org/abs/2402.08165}
\BIBentrySTDinterwordspacing

\bibitem{collini_using_2023}
\BIBentryALTinterwordspacing
L.~Collini \emph{et~al.}, ``Using {Static} {Analysis} for {Enhancing} {HLS} {Security},'' \emph{IEEE Embedded Systems Letters}, pp. 1--1, 2023. [Online]. Available: \url{https://ieeexplore.ieee.org/document/10308602/}
\BIBentrySTDinterwordspacing

\bibitem{qiu_explaining_2024}
\BIBentryALTinterwordspacing
S.~Qiu, B.~Tan, and H.~Pearce, ``Explaining {EDA} synthesis errors with {LLMs},'' Apr. 2024, arXiv:2404.07235 [cs]. Accepted to appear at the 1st IEEE Int. Workshop on LLM-Aided Design (LAD'24). [Online]. Available: \url{http://arxiv.org/abs/2404.07235}
\BIBentrySTDinterwordspacing

\bibitem{ahmad_flag_2023}
\BIBentryALTinterwordspacing
B.~Ahmad \emph{et~al.}, ``{FLAG}: {Finding} {Line} {Anomalies} (in code) with {Generative} {AI},'' Jun. 2023, arXiv:2306.12643 [cs]. [Online]. Available: \url{http://arxiv.org/abs/2306.12643}
\BIBentrySTDinterwordspacing

\bibitem{ahmad_hardware_2024}
\BIBentryALTinterwordspacing
------, ``On {Hardware} {Security} {Bug} {Code} {Fixes} {By} {Prompting} {Large} {Language} {Models},'' \emph{{IEEE} Trans. Inf. Forensics Security}, pp. 1--1, 2024. [Online]. Available: \url{https://ieeexplore.ieee.org/document/10462177}
\BIBentrySTDinterwordspacing

\bibitem{kande_security_2024}
\BIBentryALTinterwordspacing
R.~Kande \emph{et~al.}, ``({Security}) {Assertions} by {Large} {Language} {Models},'' \emph{{IEEE} Trans. Inf. Forensics Security}, 2024, preprint: https://arxiv.org/abs/2306.14027. [Online]. Available: \url{https://ieeexplore.ieee.org/document/10458667}
\BIBentrySTDinterwordspacing

\bibitem{pearce_asleep_2022}
H.~Pearce \emph{et~al.}, ``Asleep at the {Keyboard}? {Assessing} the {Security} of {GitHub} {Copilot}’s {Code} {Contributions},'' in \emph{2022 {IEEE} {Symp.} on {Security} and {Privacy} ({SP})}, May 2022, pp. 754--768.

\bibitem{opentitan_github_2024}
\BIBentryALTinterwordspacing
lowRISC, ``Opentitan,'' 2024, last accessed on 05/05/2024. [Online]. Available: \url{https://github.com/lowRISC/opentitan}
\BIBentrySTDinterwordspacing

\bibitem{noauthor_trust-huborg_nodate}
\BIBentryALTinterwordspacing
``Trust-{Hub}.org.'' [Online]. Available: \url{https://trust-hub.org/#/home}
\BIBentrySTDinterwordspacing

\bibitem{noauthor_cad4security_nodate}
\BIBentryALTinterwordspacing
``{CAD4Security} – {CAD4Security}.'' [Online]. Available: \url{http://cad4security.org/}
\BIBentrySTDinterwordspacing

\bibitem{noauthor_cad_nodate}
\BIBentryALTinterwordspacing
``\BIBforeignlanguage{en-US}{{CAD} for {Assurance} – {CAD} for {Assurance} of {Electronic} {Systems}}.'' [Online]. Available: \url{https://cadforassurance.org/}
\BIBentrySTDinterwordspacing

\bibitem{ah-kiow_investigation_2024}
\BIBentryALTinterwordspacing
J.~Ah-kiow and B.~Tan, ``An {Investigation} of {Hardware} {Security} {Bug} {Characteristics} in {Open}-{Source} {Projects},'' Feb. 2024, arXiv:2402.00684 [cs]. [Online]. Available: \url{http://arxiv.org/abs/2402.00684}
\BIBentrySTDinterwordspacing

\bibitem{fischer_hardware_2021}
M.~Fischer \emph{et~al.}, ``Hardware {Penetration} {Testing} {Knocks} {Your} {SoCs} {Off},'' \emph{IEEE Design Test}, vol.~38, no.~1, pp. 14--21, Feb. 2021.

\bibitem{noauthor_cycuity_nodate}
\BIBentryALTinterwordspacing
``\BIBforeignlanguage{en-US}{Cycuity {\textbar} {Security} assurance starts here.}'' [Online]. Available: \url{https://cycuity.com/}
\BIBentrySTDinterwordspacing

\bibitem{zhang_end--end_2018}
\BIBentryALTinterwordspacing
R.~Zhang \emph{et~al.}, ``End-to-{End} {Automated} {Exploit} {Generation} for {Validating} the {Security} of {Processor} {Designs},'' in \emph{2018 51st {Annual} {IEEE}/{ACM} {Int.} {Symp. } on {Microarchitecture} ({MICRO})}.\hskip 1em plus 0.5em minus 0.4em\relax Fukuoka: IEEE, Oct. 2018, pp. 815--827. [Online]. Available: \url{https://ieeexplore.ieee.org/document/8574588/}
\BIBentrySTDinterwordspacing

\bibitem{meng_rtl-contest_2021}
X.~Meng \emph{et~al.}, ``{RTL}-{ConTest}: {Concolic} {Testing} on {RTL} for {Detecting} {Security} {Vulnerabilities},'' \emph{{IEEE} Trans. Comput.-Aided Design Integr. Circuits Syst.}, pp. 1--1, 2021.

\bibitem{laeufer_rfuzz_2018}
\BIBentryALTinterwordspacing
K.~Laeufer \emph{et~al.}, ``\BIBforeignlanguage{en}{{RFUZZ}: coverage-directed fuzz testing of {RTL} on {FPGAs}},'' in \emph{\BIBforeignlanguage{en}{Proc. the {Int.} {Conf.} on {Computer}-{Aided} {Design}}}.\hskip 1em plus 0.5em minus 0.4em\relax San Diego California: ACM, Nov. 2018, pp. 1--8. [Online]. Available: \url{https://dl.acm.org/doi/10.1145/3240765.3240842}
\BIBentrySTDinterwordspacing

\bibitem{meza_information_2023}
\BIBentryALTinterwordspacing
A.~Meza and R.~Kastner, ``Information {Flow} {Coverage} {Metrics} for {Hardware} {Security} {Verification},'' Apr. 2023, arXiv:2304.08263 [cs]. [Online]. Available: \url{http://arxiv.org/abs/2304.08263}
\BIBentrySTDinterwordspacing

\bibitem{farzana_saif_2021}
\BIBentryALTinterwordspacing
N.~Farzana \emph{et~al.}, ``{SAIF}: {Automated} {Asset} {Identification} for {Security} {Verification} at the {Register} {Transfer} {Level},'' in \emph{2021 {IEEE} 39th {VLSI} {Test} {Symp. } ({VTS})}, Apr. 2021, pp. 1--7, iSSN: 2375-1053. [Online]. Available: \url{https://ieeexplore.ieee.org/document/9441039}
\BIBentrySTDinterwordspacing

\bibitem{zhang_transys_2020}
\BIBentryALTinterwordspacing
R.~Zhang and C.~Sturton, ``Transys: {Leveraging} {Common} {Security} {Properties} {Across} {Hardware} {Designs},'' in \emph{2020 {IEEE} {Symp. } on {Security} and {Privacy} ({SP})}.\hskip 1em plus 0.5em minus 0.4em\relax San Francisco, CA, USA: IEEE, May 2020, pp. 1713--1727. [Online]. Available: \url{https://ieeexplore.ieee.org/document/9152775/}
\BIBentrySTDinterwordspacing

\bibitem{noauthor_verilator_nodate}
\BIBentryALTinterwordspacing
``Verilator {User}’s {Guide} — {Verilator} 5.024 documentation.'' [Online]. Available: \url{https://verilator.org/guide/latest/#}
\BIBentrySTDinterwordspacing

\bibitem{wang_llms_2024}
\BIBentryALTinterwordspacing
Z.~Wang \emph{et~al.}, ``\BIBforeignlanguage{en}{{LLMs} and the {Future} of {Chip} {Design}: {Unveiling} {Security} {Risks} and {Building} {Trust}},'' May 2024, arXiv:2405.07061 [cs]. [Online]. Available: \url{http://arxiv.org/abs/2405.07061}
\BIBentrySTDinterwordspacing

\bibitem{meng_unlocking_2023}
\BIBentryALTinterwordspacing
X.~Meng \emph{et~al.}, ``Unlocking {Hardware} {Security} {Assurance}: {The} {Potential} of {LLMs},'' Aug. 2023, arXiv:2308.11042 [cs]. [Online]. Available: \url{http://arxiv.org/abs/2308.11042}
\BIBentrySTDinterwordspacing

\bibitem{trippel_bomberman_2021}
\BIBentryALTinterwordspacing
T.~Trippel \emph{et~al.}, ``Bomberman: {Defining} and {Defeating} {Hardware} {Ticking} {Timebombs} at {Design}-time,'' in \emph{2021 {IEEE} {Symp. } on {Security} and {Privacy} ({SP})}.\hskip 1em plus 0.5em minus 0.4em\relax San Francisco, CA, USA: IEEE, May 2021, pp. 970--986. [Online]. Available: \url{https://ieeexplore.ieee.org/document/9519417/}
\BIBentrySTDinterwordspacing

\bibitem{han_hardware_2019}
T.~Han, Y.~Wang, and P.~Liu, ``Hardware {Trojans} {Detection} at {Register} {Transfer} {Level} {Based} on {Machine} {Learning},'' in \emph{2019 {IEEE} {Int.} {Symp. } on {Circuits} and {Systems} ({ISCAS})}, May 2019, pp. 1--5.

\bibitem{waksman_fanci_2013}
\BIBentryALTinterwordspacing
A.~Waksman, M.~Suozzo, and S.~Sethumadhavan, ``{FANCI}: identification of stealthy malicious logic using boolean functional analysis,'' in \emph{Proc. the 2013 {ACM} {SIGSAC} Conf. on {Computer} \& communications security}.\hskip 1em plus 0.5em minus 0.4em\relax ACM, Nov. 2013, pp. 697--708. [Online]. Available: \url{https://doi.org/10.1145/2508859.2516654}
\BIBentrySTDinterwordspacing

\bibitem{gazzola_automatic_2019}
\BIBentryALTinterwordspacing
L.~Gazzola, D.~Micucci, and L.~Mariani, ``Automatic {Software} {Repair}: {A} {Survey},'' \emph{{IEEE} Trans. Softw. Eng.}, vol.~45, no.~1, pp. 34--67, Jan. 2019. [Online]. Available: \url{https://ieeexplore.ieee.org/document/8089448/}
\BIBentrySTDinterwordspacing

\bibitem{Ahmad_cirfix_2022}
\BIBentryALTinterwordspacing
H.~Ahmad, Y.~Huang, and W.~Weimer, ``Cirfix: automatically repairing defects in hardware design code,'' in \emph{ACM Conf. on Architectural Support for Programming Languages and Operating Systems}, Feb 2022, p. 990–1003. [Online]. Available: \url{https://doi.org/10.1145/3503222.3507763}
\BIBentrySTDinterwordspacing

\bibitem{laeufer_rtl-repair_2024}
\BIBentryALTinterwordspacing
K.~Laeufer \emph{et~al.}, ``\BIBforeignlanguage{en}{{RTL}-{Repair}: {Fast} {Symbolic} {Repair} of {Hardware} {Design} {Code}},'' in \emph{\BIBforeignlanguage{en}{Proc. the 29th {ACM} {Int.} {Conf.} on {Architectural} {Support} for {Programming} {Languages} and {Operating} {Systems}, {Volume} 3}}.\hskip 1em plus 0.5em minus 0.4em\relax La Jolla CA USA: ACM, Apr. 2024, pp. 867--881. [Online]. Available: \url{https://dl.acm.org/doi/10.1145/3620666.3651346}
\BIBentrySTDinterwordspacing

\end{thebibliography}

\end{document}